\documentclass[preprint,english]{revtex4}

\usepackage[T1]{fontenc}
\usepackage[latin1]{inputenc}
\usepackage{amsmath}
\usepackage{amssymb}

\makeatletter
\usepackage{graphicx}    % Include figure files
\usepackage{dcolumn}    % Align table columns on decimal point
\usepackage{bm}             % bold math
\usepackage{babel}

\begin{document}
\title{Exotic spectroscopy and decays: prospects for colliders}

\author{ J.L. Domenech-Garret$^1$}
\affiliation{$^{1}$ Departamento de F\'{\i}sica, EUITA-EI Aeron\'autica y del Espacio.\\
Univ. Polit\'{e}cnica de Madrid, 28040 Madrid, Spain. E-mail: domenech.garret@upm.es }

\date{\today}

\begin{abstract}
In addition
to well-motivated scenarios like supersymmetric particles,
the so-called exotic matter (quirky matter, hidden valley models, etc.)
can show up at the LHC and ILC,  
by exploring the spectroscopy of high mass levels and decay rates.
In this paper we use  QCD-inspired 
potential models, though without resorting
to any particular one, to calculate
level spacings of bound states and decay rates of the aforementioned
exotic matter in order to design discovery strategies.
We mainly focus on quirky matter, but our conclusions
can be extended to other similar scenarios.

\end{abstract}

\maketitle

\section{Introduction}

Since the beginning of accelerator physics,
mass spectroscopy has been playing a leading role in
the discovery of particle and resonance states,
and understanding of the fundamental interactions 
in the Standard Model (SM).
For example, the first signals of charm and bottom quarks were in fact
detected through the formation of $J/\psi$ and $\Upsilon$
bound states.

On the other hand, current colliders like the LHC, or the ILC in
a farther future,
will likely continue this discovery program beyond the SM.
It is conceivable that new (super) heavy bound states can
be formed and, contrary to e.g. the toponium system, their
basic constituents are prevented from decaying before the
binding is effective. 
The goal of this paper is to perfom a prospective
study of the spectroscopy of such exotic massive states,
by making several reasonable assumptions about the 
interacting potential among the new-physics constituents
which may differ from standard QCD.
Furthermore, we will estimate leptonic decay widths
of very heavy bound states by making specific assumptions
on the quantum numbers of constituents, although not
in a comprehensive way.

%%%%%%%%%%%%%%%%%%%%%%% 60 CARACTERES %%%%%%%%%%%%%%%%%%%%%

\subsection{Exotic scenarios}\

During the last years, minimal extensions of the SM containing
additional heavy particles charged under a new unbroken non-Abelian
gauge group $G_v$ with  fermions \emph{Q,$\bar{Q}$} have been proposed
under the general name of $\lq\lq$hidden valley'' models 
\cite{Strassler:2006im}, which is a very general scenario containing such heavy particles 
but as well new sectors of lightweight particles to be observed. 
In these models all SM particles would be neutral
under such the new $G_v$ group, while new particles 
charged under $G_v$ but neutral under the SM group would
show up if large energy scales are probed. Higher dimension operators, 
induced e.g. by  a $Z'$ or a loop involving heavy particles carrying both $G_{SM}$ and
$G_v$ charges, should connect both SM and new physics sectors 
through rather weak interactions.

In particular, if the $G_v$ group corresponds to $SU(3)$, the 
fermions in the fundamental representation have been recently named 
as $\lq\lq$quirks" \cite{Luty} or iquarks \cite{cheung}. 
\cite{comm1} Actually this theory can be viewed as a 
certain limit of QCD where light quarks are removed and the typical scale
$\Lambda$ where the new interaction becomes strong is much smaller
than the heavy flavor masses. More generally, such kind of scenario can be
put in correspondence with a class of $\lq\lq$hidden valley'' models
as pointed out in \cite{Strassler:2006im}. In these models, quirks are defined to 
be new massive fermions transforming under both $G_{SM}$ and a general 
( not only $SU(3)\ $) non-abelian gauge group $G_v$. 

It has been considered so far in the
literature that $\Lambda$ is smaller than $\Lambda_{QCD}$, but as well 
the particular case of hidden valleys with quirks in which  $\Lambda$ 
is greater than $\Lambda_{QCD}$ has been studied in the literature 
\cite{Strassler:2006im},\cite{Juknevich},\cite{Martin}. The name of 
{\em infracolour} is used to design of the new gluonic degrees of freedom
when $\Lambda$ is much smaller than the quirk mass; in this work 
we will refer to it as i-colour (i-QCD) in correspondence with the name quirk.

\subsection{Exotic long-lived bound states}\

In this section we will focus on the bound states of quirks. The phenomenology
of such bound states has first studied in \cite{Luty}, and later an analysis on 
the spectroscopy of these systems was done by the authors of \cite{kribs}.\\  
 
It is well-known that the large mass of the top quark in the SM prevents
toponium to be formed since the constituents quarks would
decay away too fast. A criterion for existence of
such bound states is that the binding energy should be
larger than the total decay width. 
For heavy onia states beyond the SM, however, the situation could be
different. In particular, in the case of quirks \cite{Luty},\cite{okun},\cite{kribs},
its decay is prevented from the conservation of a quantum number.
Regarding the dynamics of these quirks, according to \cite{Luty}, one can distinguish among three
possible energy scales: the first one is the  $100\ \mathrm{eV}\ \lesssim\ \Lambda \lesssim\ 
\mathrm{keV}$ range, where the quirk strings are macroscopic; the  mesoscopic strings 
can be find at the range   $\mathrm{keV}\ \lesssim\ \Lambda \lesssim\ \mathrm{MeV}$
(which is large compared to the atomic scales); and  finally we find the microscopic scale at  
$\mathrm{MeV}\ \lesssim\ \Lambda \ \lesssim\ m_Q$.\\

On the other hand, assuming that the scale of i-colour is below
the weak scale, bound states of the new sector are kinematically
accessible to present and future collider experiments. However, since
the SM particles are uncharged under i-colour, (quirk) loops
would be required to couple both sectors leading to highly
supressed production rates. Moreover, from reference \cite{Luty}, quirks 
are defined to be charged under (some) gauge groups of the SM (for example, charged under 
the electroweak interaction) quirks could be pair produced through electroweak
processes.\\

Likewise, in Ref.\cite{cheung} quirks are considered as vectors
with respect to the electroweak gauge group without carrying
QCD colour, but carrying i-colour charge.
Therefore notice that there is no Yukawa coupling between the
Higgs boson and quirks. Thus we discard the possible binding force 
which has been postulated for ultraheavy quarkonium taking over
gluon-exchange. At this point, for the sake of clarity, 
it must be stressed that, taking simple assumptions, through 
this work we will focus on the case of uncoloured quirks.\\  

In this scenario, quirks can be copiously pair-produced at the LHC
not through QCD interactions but via electromagnetic and 
electroweak interactions. As quirks would be long-lived particles 
as compared to the collider/detection time scale, different detection 
strategies can be undertaken according to the possible aforementioned 
micro, meso or macroscopic regimes.\\

Finally, notice that possible
quirky signals of folded supersymmetry in colliders have
been studied in \cite{Burdman:2008ek}, focusing on
the scalar quirks ({\em squirks}). Contrary to usual supersymmetric
partners of quarks, squirks (choosing a simple scenario) are expected to be uncolored, but
instead charged under a new confining group, equivalent to i-QCD
as introduced above. The study of spectroscopy performed in this paper 
is actually not sensitive to possible scalar nature of new fermions, 
and the main lines could be applicable
to squirkonium as well.

\subsection{Exotic phenomenology}\

According to \cite{Martin}, when the quirk pair is produced an excited  bound state can be formed
with invariant mass given approximately by the total center of momentum energy of the hard partonic 
scattering giving raise to the pair. In the microscopic regime this bound state would loose energy 
by emitting i-glueballs and bremsstrahlung  towards low-lying states. 
Once they loose most of their kinetic energy these bound states (to be dubbed Quirkonium) could decay 
via electroweak interaction. However, it is also possible other scenario in which the neutral and 
colourless quirk pair might have a prompt annihilation before they can loose energy  enough to form the 
low lying quirkonium state. However, due to the non perturvative nature of the mechanisms it is difficult to estimate 
in advance the proportion of the events falling in each scenario,and therefore the possibility to detect 
these low lying states can not be discarded. Through this work, 
in particular we will focus on neutral bound states which can decay, 
e.g. to final-state dileptons, providing a clean signature even 
admits a huge hadronic background as at the LHC experiments.\\

New particles with a mass of up to several hundred GeV can be
pair copiously produced at the LHC. One expects that quirks will be 
in general produced with kinetic energy quite larger than $\Lambda$. 
A significant fraction of this energy should be lost by emission of photons and
i-glueballs prior to pair-annihilation. 

The two quirks will fly off back-to-back, developing
a i-QCD string or flux tube. In usual QCD with light
matter the string is broken up promptly by creating
light quark-antiquark pair; in i-QCD this mechanism
is practically absent.  The two heavy ends
of the string would continue to move apart, eventually
stopping once all the kinetic energy was stored in the string.
The quirks would be then pulled together by the string
beginning an oscillatory motion.

Most examples of late-decaying particles that have been addressed
in the literature yield missing energy, while quirks would annihilate
into visible energy in most modes. Besides, as explained in \cite{Luty} 
and \cite{okun}, only i-colour singlet  states could be observed.

As commented before, in a optimistic scenario, the excited bound state 
will emit i-glueballs and bremsstrahlung, towards low-lying states;
then they would annihilate into a hard final state:
di-lepton, di-jet, or di-photon.\\

If we want to investigate whether or not it is possible to disentangle different state levels
under the assumption of a given (large) quirk mass and a specific
form of new i-colour interaction, then  we could  focus on a dimuon signal from the annihilation 
of a {\em narrow} resonance, since it is the most promising channel \cite{Martin}; 
it then becomes crucial if the level spacing of different $S$-wave
states is enough to  the foreseen mass resolution based on invariant mass 
reconstruction from a dimuon system.
 
According to the reference \cite{Martin} the detection of bound states 
at hadron colliders is reliable because the signal production is strong and peaked in invariant mass, 
and the dominant  background is electroweak and diffuse. On the other hand, dimuon backgrounds from sources 
other than Drell-Yan  can be suppressed by requiring no extra hard jets or missing energy. 
Besides, at the LHC, trigger and detector  efficiencies are expected to be very high for high-mass 
dimuon events.\\

Concerning other channels, it is expected that Quirkonium annihilation 
into a electron pair could be also a useful
signal since the invariant mass peak is expected to 
yield a similar peak to dimuon but smaller and wider due to 
detector effects.

On the other hand, in the squirkonium
case, according to the reference \cite{Harnik}
the  radiative decay (by soft radiation) from the highly excited states
to the ground state  can be ultimately  detected by means of unclustered 
soft fotons in the uncolored case. Also, in ref. \cite{Burdman:2008ek} it is 
pointed out the possibility to use the invariant mass peak of  $W+$photon, 
since this channel dominates the squirkonium  annhilation at or 
near the ground state. If necessary, all these signals could 
aid to distinguish among different states.\\

Focusing again on the dimuon signal, as it can be seen from  ATLAS and CMS  
reports \cite{ATLAS:1999fq},\cite{CMS}
one can consider a $2\%$ accuracy for the transverse momentum of muons even at high momentum. Since 
the dimuon invariant mass should coincide to a good approximation, for small (pseudo)rapidities,
 with the transverse momentum error, $\delta(M_{\mu\mu})\ \simeq\ 0.02\ \times\ M_{Q\bar{Q}}$ (since 
$p_t\ \simeq\ M_{Q\bar{Q}}/2$);   letting $M_{Q\bar{Q}}$ vary along the range $[100,1000]$ GeV,
the mass resolution should roughly take the values along the interval $[2,20]$ GeV.  

%\newpage
\subsection{Model settings}\

Hereafter we restrict our analysis to the range of $\Lambda$ given by
$\Lambda << m_Q$ but in the microscopic regime, namely
 \begin{equation}
 \mathrm{few}\ \mathrm{MeV}\ \lesssim\ \Lambda \ \lesssim\ \mathrm{few}\  
\mathrm{GeV} 
 \end{equation}

As is well-known long ago, a  non-relativistic treatment 
of the potential for conventional heavy quarkonium has
proved to be suitable on account of the asymptotic freedom of QCD.
Moreover, one distinguishes between 
short- and long-distance dynamics of constituents in the bound state
leading to an effective (static)
potential of the type:
\begin{equation}
\label{eq:potential}
V(r)\ =\ V_S(r)\ +\ V_L(r) 
\end{equation}
In particular we will write
\begin{equation}
\label{eq:potential2}
V(r)=\ -\ \frac{A}{r^{\mu}}\ +\ B\ r^{\nu}
\end{equation}
where the first term with $\mu=1$ would correspond to a Coulombic
interaction, and the second one with $\nu=1$ to a linear confining
interaction. 

In this work we will consider firstly a Coulomb
plus Linear potential (CpL)  with $\mu=1$ and $\nu=1$; later
we will use a more general Coulomb plus Power Law  potential (CpP)
with $\mu=1$ and $\nu=0.5,1.5$ as
tentative possibilities.  The motivation for the insertion of these
power law potentials comes up from the clasical 
studies of Quarkonia (see Refs. \cite{flamm}, \cite{Rosner}, \cite{Eichten}.) 
in which are considered Coulomb like, linear and Cornell potentials, but as well 
the power law potentials are taken into account in order to cover possible deviations from them. 
In this way and focusing on the case of Quirkonium, in reference \cite{Luty} 
a pure linear potential is taken into account, in  reference \cite{kribs} a 
Coulomb-like potential were considered. Therefore tracking the same philosophy  
than in the Quarkonia case possible deviations are  also considered within this study.

Moreover, the interaction accounting for the above static potential can
be parameterized by the fermion (quirk) mass
$m_Q$, where $100\ \mathrm{GeV} \lesssim m_Q \lesssim  \mathrm{TeV}$
and an additional $SU(N_{IC})$ gauge  coupling ($N_{IC}$ stands for the
i-colour number) which can be related with a i-colour scale $\Lambda$.

In this work we will specify $V(r)$ in Eq.(\ref{eq:potential2}) as
\begin{equation}
\label{eq:potential3}
V(r)= \sigma\  r^{\nu} - \frac{C\  \alpha'}{r}
\end{equation}
to be interpreted as 
CpL ($\nu=1$) and CpP ($\nu=0.5,1.5$) potentials.
Here $\sigma$ corresponds to the i-color string tension
and we have introduced a i-color coupling
$\alpha'$, alongside a group theory factor $C$, in
close analogy to QCD potential models; hence,
making a simple assumption, such group factor is taken as  a mirror from  QCD potentials
($C=4/3$) and included into the infracolour coupling: i.e., in calculations we set
$\alpha_{Icolour}= C \alpha'$. Of course, other numerical choices for $C$ can be done, but 
$\alpha'$ depends on the $\Lambda$ scale which is actually uncertain as we shall discuss later.\\

On the other hand, in analogy to conventional QCD-inspired potential models,
the i-colour string tension
can be interpreted as a linear energy
density ($E/L$), where $E \sim \Lambda$
and $L\sim \Lambda^{-1}$. Hence the relation 
$\sigma \sim   \Lambda^2$ is expected to remain (approximately) valid,
likewise the equivalent QCD expression $\sigma_{s} \sim \Lambda_{QCD}^2$
(also derived from lattice calculations \cite{donahue}), and finally

\begin{equation}\label{sigmalambda}
\sigma \propto \ \biggl[ \frac{\Lambda}{\Lambda_{QCD}}\ \biggr]^2\ \sigma_{s}
\end{equation}

In other words, a proportionality depending on the respective
$\Lambda$ and $\Lambda_{QCD}$ between both string tensions could be
expected from the above arguments. Basically, Eq.(\ref{sigmalambda})
implies that $\sigma$ and $\Lambda$ parameters are not independent of
each other. For the sake of simplicity, the unknown proportionality
factor will be set equal to unity, so that by fixing $\Lambda$ one
gets $\sigma$ (for given $\Lambda_{QCD}$ and $\sigma_s$ values).

Focusing on the  $\Lambda$  scale, in this work in principle it corresponds to   
the microscopic $\Lambda'$ scale depicted in Ref.\cite{cheung}. 
Numerically speaking, as it will be seen, the values of 
$\Lambda$ were taken below and above of the QCD scale in a bandwidth; i.e.
$\Lambda = k \Lambda_{QCD}$ with $k=(0.1, 0.4, 1, 10)$ to take into account the uncertainity 
about this quantity. In this way, the equation(\ref{sigmalambda}) can be regarded 
as a comparison between the strength of the linear potential in both sectors 
$SU(3)_{QCD}$ and the new gauge group $SU(3)_{Icolour}$, and it is intended to be an 
ansatz or a hint to determine numerically a proportionality between $\sigma_{s}$ and 
$\sigma$, in which subsequently  the numerical uncertainity about the proportionality 
factor is diluted, taking into account the lack of knowledge about $\Lambda$. 
Besides, this comparison between diferent theories can be viewed to some extent 
in a similar way than in classical physics, in which 
the strength of gravitational and the electrostatic forces are compared.\\

Concerning the i-colour coupling constant, $\alpha'$, it 
would be related with $\Lambda$; as we are dealing with a
 non-Abelian i-colour binding force it implies  
$\alpha'$ is scale dependent. We will compute
 $\alpha'$ value at the running scale $Q=2m_Q$
according to \cite{cheung}: 
 \begin{equation}\label{eq:alpha_s}
 \alpha'(Q) =  \frac{12\ \pi}{(11\ N_{IC}-2n_Q)\
\ln{\biggl( \frac{Q^2}{\Lambda^2}\biggr)} }
 \end{equation}
where $N_{IC}$ is the i-colour number, and $n_Q$ the number of quirk generation at the running scale. From 
Eq.(\ref{sigmalambda}) and Eq.(\ref{eq:alpha_s}), one can see
that both parameters $\sigma$ and $\alpha'$ are  depending on $\Lambda$, so that they are not independent quantities. Nonetheless, all those constraints
have to be taken with a grain of salt and one should consider as well
values deviating from those given in Eqs. (\ref{sigmalambda} - \ref{eq:alpha_s}), as we will see later.

In case of more quirk generations, additional active quirk should be taken into
account at different energy scale thresholds. 
Nevertheless, for the sake of simplicity, and in view of still many unknowns
in the different models, we will assume $N_{IC}=3$ and  $n_Q=1$ 
throughout this work. \\

 As previously mentioned, we consider that the quirk mass lies in the 
range $100\ \mathrm{GeV} \lesssim\ m_Q \lesssim\  \mathrm{TeV}$. 
Therefore one can reasonably expect that the bound system indeed meets
a truly non-relativistic regime, i.e. the
relative quirk velocity $v$ in the center of mass frame 
being substantially smaller than the value for bottomonium
($v^2 \simeq 10^{-1}$). Focusing on quirkonium, a formal derivation of 
such non-relativistic limit from the relativistic degrees of freedom can 
be found in \cite{kribs}; besides according to Reference \cite{Luty} and \cite{Harnik},  
the bound state is formed in a highly excited state then it decays to the lower states 
loosing the main part of its kinetic energy. Therefore it is expected that in  the lower 
levels near to the ground state this kinetic energy could be low enough to assume a 
non-relativistic aproximation.  Also, as we shall see later, the numerical results obtained for the expected 
quirk velocity $<v^2>$ in the CoM frame justify this approximation. 

\section{Prospective spectroscopy of exotic states} 

Since we are interested to perform spectroscopy for very heavy  
non-relativistic bound  states, the Schr\"{o}dinger radial equation
must be solved: in a analytical way it could be done by means of a expansion 
of the quirkonium wave function in  a complete basis; nevertheless here we will choose
to solve it numerically, and therefore one should expect 
that the method to get the resulting mass spectroscopy followed in the \emph{QQ-onia} 
package \cite{JL} based on the resolution of the Schr\"{o}dinger radial equation
using the Numerov $\textit{0}(h^6)$ technique, should work appropriately
for our analysis of Quirkonium. However, we are confronted here to the lack 
of experimental data to set the ground state of quirkonium, in sharp contrast 
with, e.g., the bottomonium or charmonium systems. 
Nevertheless, let us stress that in this work we are here mainly 
interested in estimating the mass spacing between different state levels
rather than their absolute values.\\

As commented in the Introduction, new interactions
and particles can form very massive bound states. In this section
we show the results for quirk $(Q\bar{Q})$ bound state
system by sweeping through the scale
range, $few\ \mathrm{MeV}\ \lesssim\ \Lambda \ \lesssim\ few\  GeV$ 
characterizing the i-colour force.

 First we will look at the results using  
a Coulomb plus Linear potential (CpL)  (with $\mu=1$ and $\nu=1$);
later using a Coulomb plus Power Law potential (CpP) (with $\mu=1$
and $\nu=0.5,1.5$). Concerning the quirk mass, first we use  $m_Q=
100$ GeV, and later $m_Q= 500$ GeV as representative values
(nevertheless  in some calculations
 we will take additional values). The energy
levels, $E_n$ shown in Tables correspond to $ M_{nl} =\ 2 m_Q + E_{nl}$
where $M_{nl}$ (or $M$) is the quirkonium mass level. In all cases, 
we set the ground state to be $E_{1S}= 0$.
Besides as  relevant calculations we  will display also the  squared
radial wave function at the origin (WFO) (or their derivatives), the
size of each quirkonium level, and the mean value of the  relative 
quirk velocity $v$ in their center of mass frame, since it is used
in some calculations \cite{Luty}; besides, the obtained velocities  
will check the non-relativistic approximation. 
\newpage

\subsection{\textbf{CpL potential}}\

Let us start by considering the CpL potential. All Tables cited
in this and successive sections can be found in Appendix.

\subsubsection{$m_Q= 100\ \mathrm{GeV}$}\

Results for the quirkonium spectrum with $m_Q=100$ GeV and
$\Lambda=\ 0.1\ \Lambda_{QCD}= 25$ MeV are shown in Table \ref{table:1}.
 The corresponding parameters are $\sigma=0.0018\ \mathrm{GeV}^2$ and   
$\alpha'(Q=2m_Q)=0.068$. Concerning the mean radius, for the ground state we find a size similar to the Bohr radius $r_B \sim (m\alpha')^{-1}$,
 and increases for higher states as expected up to $ \sim 1 fm$; moreover, with these
parameters we can found ($ 8S $) states with sizes beyond $2 fm$, which is in accordance with $\sim \Lambda^{-1}$.
The quirk velocity in the CoM frame $\langle v^2 \rangle \approx 10^{-4}$ 
slowly increasing with the $n$ and $l$ quantum numbers; these low $v$ values
plainly justify the non relativistic regime resulting from the
QQ-onia package.\\

  We provide the squared WFO and derivatives divided
by powers of the quirk mass obtained in our calculations,
following the same behaviour with $n$ and $l$ as found
in standard quarkonium (see for instance \cite{JL} and references therein). From the ground state WFO value
we realize that $1S$ state follows mainly a Coulombic  behaviour \cite{comm2}. This is not the case for higher resonances, for the P states case we find that a  Coulombic (derivative) WFO  underestimates the numerical value obtained from this potential.

Let us stress that the energy level spacing (notably
between S-wave resonances, of order of tens of MeV) 
would not permit the experimental discrimination by using the dimuon annihilation channel (and 
likely any other else).

Table \ref{table:2}  shows the results for $\Lambda=\ 0.4 \cdot \Lambda_{QCD}= 100\ 
\mathrm{MeV}$ ($\sigma=0.029\  \mathrm{GeV}^2$ and
$\alpha'(Q=2m_Q)=0.08$). The $Q\bar{Q}$ level spacings 
turn out to be somewhat larger than in the previous case
but still not enough to permit experimental discrimination.
Something similar can be expected for 
$\Lambda=\Lambda_{QCD}= 250$ MeV as can be seen from Table \ref{table:3}, with ($\sigma=0.18\ \mathrm{GeV}^2$ and $\alpha'(Q=2m_Q)=
0.091$).  Here we find lower values for sizes of resonances with respect to previous case  (as expected since $\Lambda$ increases). Besides, we observe a WFO value for the ground state somewhat greater than the expected for a Coulombic  behaviour.

The results shown in Table \ref{table:4} (appendix ) corresponds to the microscopic scale
$\Lambda=\ 10\ \Lambda_{QCD}$. Here $\alpha'(Q=2m_Q) =\ 0.139$,
and  $\sigma = 18\  \mathrm{GeV}^2$. 
The string tension turns to be much stronger than
in the QCD case. The  energy levels reach the GeV scale and the WFO
grow to the $\sim 10^3$ GeV$^3$ values; according with  previous
trend the corresponding derivatives  are growing also. The  WFO value for the lowest state is $\sim 3$ times 
greater than the expected for a Coulombic  behaviour. Concerning the level
spacing this case is interesting since values among $S$-wave states turns out to be of order of $\sim 2$ GeV, likely enough to be disentangled.

\subsubsection{$m_Q = 500\   GeV$}\

We now set the quirk mass equal to 500 GeV, so 
quirkonium mass is of order of the TeV
scale. In Table \ref{table:5} the $Q\bar{Q}$ spectrum is shown for
 $\Lambda=\ \Lambda_{QCD}$ ($\sigma=0.18$ GeV~$^2$  and 
$\alpha'(Q=2m_Q)=0.073$).
In Table \ref{table:6} we show the results for $\Lambda=\ 10\ \Lambda_{QCD}$. Here
$\alpha'(Q=2m_Q) =\ 0.102$, and  $\sigma = 18$ GeV~$^2$. Again, as in the $m_Q = 100$ GeV case at this scale, the level
spacing  among $S$-wave states could be enough to distinguish experimentally these levels.\\

\subsection{CpP potential.}\

Let us now give $\nu$ values in the long-range term
of Eq.(\ref{eq:potential3}) different from unity. As
in QCD, a larger (smaller) $\nu$ leads to stronger (weaker)
long-distance interaction. The general trends are similar
to the CpL case seen in the previous section.

\subsubsection{\textbf{$\nu=0.5$}}\

Tables \ref{table:7} and \ref{table:8} show the $Q\bar{Q}$ spectrum for $m_Q =
100\  \mathrm{GeV}$ and $m_Q = 500\ \mathrm{GeV}$ 
respectively for $\Lambda = \Lambda_{QCD}$. 
As we can see the sizes of bound states are similar
to the bottomonium case \cite{JL},\cite{Kinoshita}.  Tables \ref{table:9} and 
\ref{table:10} (appendix) provides again 
the corresponding $Q\bar{Q}$ spectrum for
$m_Q = 100\   GeV$  and $m_Q = 500\   GeV$, but this time having set
$\Lambda=\ 10 \cdot \Lambda_{QCD}$. Here we observe  values of  WFO  for the ground state  
similar to the ones   expected for a Coulombic  behaviour; 
however higher resonances do not  behave in this manner.

\subsubsection{\textbf{$\nu=1.5$}}\

To cover possible deviations from linear behaviour of the long distance part of the potential  we analyze
the CpP  potential  setting  $\nu=1.5$. 
Tables \ref{table:11} and \ref{table:12} (appendix) show the $Q\bar{Q}$ spectrum for with
$\Lambda=\ \Lambda_{QCD}$ for $m_Q = 100$ GeV  and $m_Q = 500$ GeV 
respectively. Tables \ref{table:13} and \ref{table:14} display the
corresponding results for  the  $Q\bar{Q}$ spectrum for $m_Q = 100$ GeV  
and $m_Q = 500$ GeV with $\Lambda=\ 10\ \Lambda_{QCD}$.\\

In order to compare the effect of the above mentioned  potentials, in Figure 1 we plot the $nS$ level spacings $\Delta_{nS-1S}=M(nS)-M(1S)$ of quirkonium found with the CpL and CpP potentials ($\nu=1,0.5,1.5$ respectively) for different $m_Q$ and $\Lambda$ values. As far as we are interested in 
disentangling peaks of $S$-wave resonances, it becomes apparent that this would be only possible
in some cases (i.e. $\Lambda=\ 10\ \Lambda_{QCD}$) where the level spacing is ${\cal O}(1)$ GeV
or larger.

\subsection{\textbf{Other possible contributions from short distance potential}}\

Finally, to take into account other contributions which could be entangled in  the short distance part of the potential, we consider higher (non perturbative) $\bar{\alpha'}$ effective values. In order to consider this scenario, we  do not use  the Eq.(\ref{eq:alpha_s}) for $\alpha'$ but we take it as a free parameter. On the other hand we keep  the  explicit dependence of $\Lambda$ ( Eq.(\ref{sigmalambda}) ) in  $\sigma$.  In this case  we  
also increase the $\Lambda$ values from $\Lambda=\ 10 \cdot \Lambda_{QCD}$  up to $\Lambda=\ 40 \cdot \Lambda_{QCD}$.
 
By using \emph{QQ-onia} code we find the results with $m_Q = 500$ GeV  for the mass level spacing  $\Delta_{2S-1S}=M(2S)-M(1S)$ which are shown in Table \ref{table:15} (Appendix). As we can see from these situations,  we find  separation between  levels   tens of GeV, thus, in principle, we should be able to discriminate at least between these resonances.

\subsection{WFO vs. quirk masses.}\

Next let us analyze the WFO dependence w.r.t. the quirk mass
using the above explained CpP potentials. Here we focus on the $1S$
ground level, by taking quirk mass values from $m_Q= 100\ GeV$ up to $500\
GeV$. Again we will take $\Lambda=\ \Lambda_{QCD}, \ 10\
\Lambda_{QCD}$ for each potential. For intermediate $m_Q$ values
$\sigma$ does not change w.r.t. the mass values, However here
$\alpha'(Q=2m_Q)$ changes for each case according to Eq.(\ref{eq:alpha_s}).
Figure 2 displays the obtained results.

\section{Quirkonium decay}\

Once computed the squared WFO, we can 
evaluate numerically the partial decay widths of neutral ($^3S_1$) 
quirkonium ($\psi_{Q\bar{Q}}$) to different final states.
All of them are proportional to the
ratio $|R_{S}(0)|^2/M^2$, where $M$ is the quirkonium mass.
Subsequently we make estimates of the 
respective branching ratios ($BR$).\\

We will follow a similar treatment as the
authors of \cite{cheung},\cite{barger} who considered the
following  $\psi_{Q\bar{Q}}$ decay modes:

\begin{itemize}
    \item Decay to Standard Model fermion pairs 
($f\bar{f}\equiv$leptons  and  quarks)
    %$\Gamma(\psi_{Q\bar{Q}} \longrightarrow f\bar{f})$
\begin{equation}\label{decayff}
\Gamma(\psi_{Q\bar{Q}} \longrightarrow f\bar{f})=
F^{f\bar{f}}_1(N_{IC}, R_i, e_Q, SM) \frac{|R_{S}(0)|^2}{M^2}
 \end{equation}
where $F^{f\bar{f}}_1(N_{IC}, R_i, e_Q, SM)$ stands for  functions
containing the i-colour number, squared mass ratios $R_i=M^2_{i}/M^2
(i =f, Z)$ , the quirk electric charge $e_Q$. The $SM$ label means
that those SM parameters involved in this
calculation parameters are included.

    \item Decay to a $W^{\pm}$ pair
    %$\Gamma(\psi_{Q\bar{Q}} \longrightarrow W^{\pm})$
 \begin{equation} \label{decayww}
 \Gamma(\psi_{Q\bar{Q}} \longrightarrow W^{+}W^{-})= F_2(N_{IC}, R_i,
SM) \frac{|R_{S}(0)|^2}{M^2}
 \end{equation}
Where $F_2(N_{IC}, R_i, e_Q, SM)$ stands for a function entangling
the i-colour number, squared mass ratios $R_i=M^2_{i}/M^2 (i =W,\
Z,\ m_Q)$,  and $SM$ parameters.

    \item Decay to i-gluons ($g'$). Quirks couple to the i-gluon field
of the $SU(N_{IC})$ with coupling strength $g_s'=\sqrt{4\pi\alpha_s'}$,
where $\alpha_s'(Q^2)$ is given by Eq.(\ref{eq:alpha_s}).
\begin{equation}\label{decay3g}
  \Gamma(\psi_{Q\bar{Q}} \longrightarrow g' g' g' )=
F_3(N^2_{IC}, \alpha'^{3}) \frac{|R_{S}(0)|^2}{M^2}
\end{equation}

Here, $F_3(N^{2}_{IC}, \alpha'^{3})$ is a function of the i-colour
number and the i-colour $\alpha'$ coupling.
\begin{equation}\label{decayp2g}
    \Gamma(\psi_{Q\bar{Q}} \longrightarrow \gamma\ g' g')= F_4(N_{IC}, e^2_Q, \alpha'^{2}) \frac{|R_{S}(0)|^2}{M^2}
\end{equation}
\begin{equation}\label{decayz2g}
    \Gamma(\psi_{Q\bar{Q}} \longrightarrow Z\ g' g')= F_5(N_{IC}, e^2_Q, \alpha'^{2}, SM)\frac{|R_{S}(0)|^2}{M^2}
\end{equation}

\end{itemize}

To make the reading easy, the explicit form of $F_i (i=1,..., 5)$ coefficients can be found in references \cite{cheung},\cite{barger}.

\subsection{Numerical results}\

Once set the numerical values of parameters in the above expressions,
the $|R_{S}(0)|^2$ values
from Tables \ref{table:1} to \ref{table:14} (appendix) allow one to compute the decay widths of
$\psi_{Q\bar{Q}}(1S)$ to $SM$ quarks ($q\bar{q} \equiv\  u\bar{u},\
d\bar{d},\ s\bar{s},\ c\bar{c},\ b\bar{b}$, [$t\bar{t}$ if
above the threshold]), leptons ($\ell\bar{\ell} \equiv  e^{\pm},\
\mu^{\pm},\  \tau^{\pm}$), and other boson decays ($W^{\pm},\
3g',\ \gamma 2g',\ Z2g'$). The results are shown in Table \ref{table:16} 
for CpL and CpP (with $\nu=0.5, 1.5$) potential using  $m_Q=100\ GeV$ and
$m_Q=500\ GeV$ at the above
considered scales.

In all cases the decay mode to $SM$ quarks is the dominant channel.
Decay to leptons shares roughly with a $33\%$ for electron, muon and
$\tau$ pair respectively. As we can see, for $\Lambda= \Lambda_{QCD}$ case if we take into account
only the $1S$ decay, the total width is quite narrow $\sim$KeV, but we
find similar values than in the heavy quarkonia case \cite{pdg}. For $\Lambda= 10 \Lambda_{QCD}$ case the total width
increases roughly one order of magnitude. Nevertheless, if necessary, this analysis could be improved by adding upper $nS$ levels
contributions\footnote{It could be also considered together with the
$^1S_0$ decays, which are proportional to $|R_{S}(0)|^2/M^2$.}:
for instance if we compute the whole $nS$ contribution using the
$m_q=500 GeV;\ \Lambda= 10 \Lambda_{QCD}$ case, with a CpL potential
we find a total width $\approx 1.7$ times  the
$1S$ total width; using CpP $\nu=1.5$ and $\nu=0.5$ potentials we find a factor $\approx 1.5$ and  $\approx 2.$ respectively.\\

Concerning $P$-wave resonances ($l\neq 0$), 
the corresponding widths satisfy 
\begin{equation}
\Gamma_{nP} \propto \frac{|R'_{S}(0)|^2}{M^4}
\end{equation}
so that, those contributions are suppressed with respect to the $nS$
decays by a ($\textsl{D}$) factor 
\[
\textsl{D}= \frac{1}{ M^2}\ \frac{|R'_{P}(0)|^2}{|R_{S}(0)|^2}  
\]

Taking $n = 1$ values  in the CpL case from  Tables  , we find  $\textsl{D}\sim
[10^{-5},10^{-4}]$ in the $m_Q = 100\ GeV$ case and $\textsl{D}\sim
[10^{-6},10^{-5}]$ for $m_Q= 500\ GeV$.
%\newpage
Regarding the dependence of the BRs on the quirk
mass ($BR$ are independent of the ratio $|R_{S}(0)|^2/M^2$, but
$M$ enters also thorugh the functions $F_i$):  in the range of
interest $100\leq m_Q \leq 500 GeV$ we find variations on  the
different $BR$ less than a $1\%$.

 We can also check the $BR$ variations
with the i-colour number $N_{IC}$: by replacing for instance in the
above expressions $N_{IC}=6\leftrightarrow N_{IC}=3 \leftrightarrow
N_{IC}=1$, the $BR$ to bosons varies mainly $33\% \leftrightarrow
17\% \leftrightarrow 5\%$ and the corresponding $BR$ to $SM$ quarks
$55 \%\leftrightarrow 68\% \leftrightarrow 78\%$ ($BR$ to
leptons $12 \%\leftrightarrow 15\% \leftrightarrow 17\%$ respectively).\\

\section{Conclusions}

Spectroscopy of exotic states might play a fundamental
role in the discovery strategy of new physics at the LHC and ILC.
In this paper we have focused on a simple extension of the SM, 
when a new $SU(N_i)$
gauge group is added to the SM. The new interaction and new associated
fermions have been dubbed i-color, quirks respectively. We assume
that quirks are colorless, but otherwise carry SM quantum numbers,
thereby coupling to gauge $W^{\pm},Z$ and $\gamma$ bosons. 

Quirks can bind forming very peculiar structures reminding. In this work we have focused on neutral $Q\bar{Q}$ states
called quirkonium, when the states are microscopic. We have performed a prospective study of quirkonium spectroscopy 
by employing a Coulomb plus Linear and Coulomb plus Power Law 
potentials  as representative possibilities with  parameters according
to i-QCD requirements, as well as  other effective  contributions  to analyze their impact.  

Taking into account the
wide range where the $Q\bar{Q}$ bound state might be found, we have
chosen the scale range, MeV $\lesssim \Lambda \lesssim$ GeV
with different i-colour $\Lambda$ scales and quirk masses, finding
sizes of several $Q\bar{Q}_{nl}$ resonances  and their squared WFO values of states (or derivatives for
$l\neq 0$). We also extracted  the level spacing among resonances  using different $\Lambda$ scenarios to determine whether or not it would be possible to discriminate different state levels. We also have computed  total and partial decay widths.

%%%%%%%%%%%%%%%%%%%%%%% 60 CARACTERES %%%%%%%%%%%%%%%%%%%%%

\begin{acknowledgments}
I am grateful to Miguel Angel Sanchis-Lozano for calling  my attention to quirkonium systems, to point out all the  details concerning detection by means of the dimuon channel, and many discussions.
\end{acknowledgments}

%\newpage

\newpage

%%%%%% FIGURES %%%%%

% Figure 1
\newpage
\vspace{2.0cm}
\begin{figure}
\includegraphics[width=8cm]{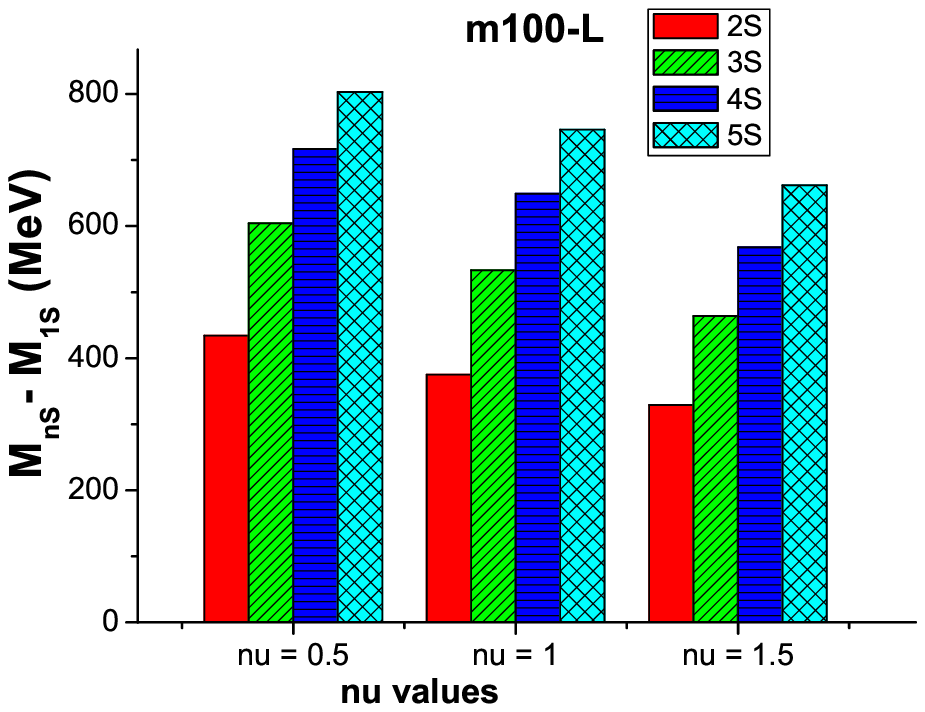} 
\includegraphics[width=8cm]{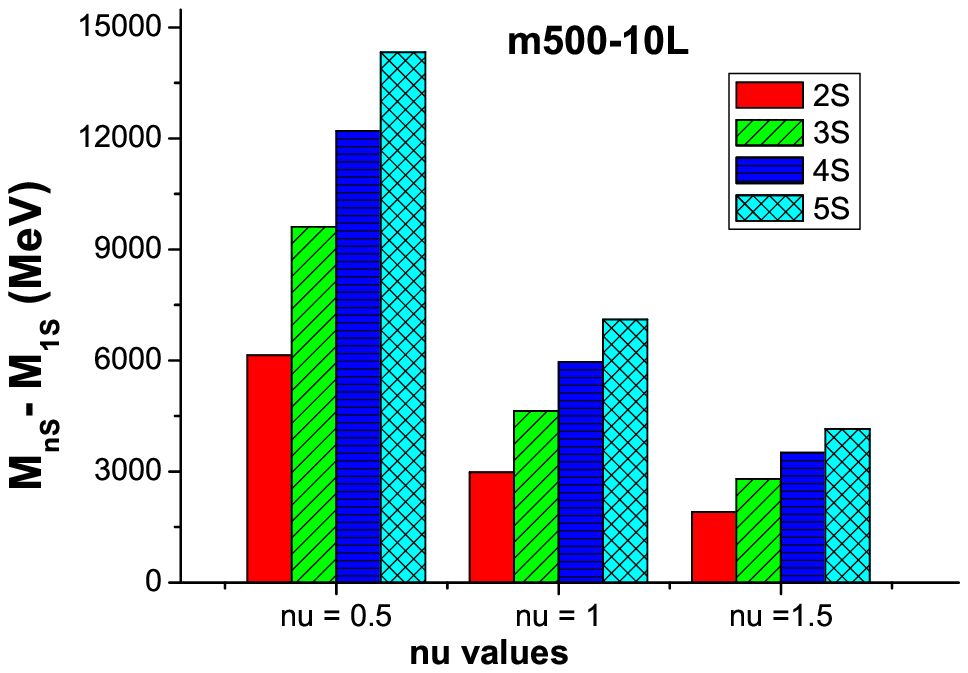}
\caption{\label{fig:1} Comparative plot of the $nS$ energy levels w.r.t. 
the $1S$ state, taken as the ground level, found with 
the CpL and CpP potentials ($\nu=1,0.5,1.5$ respectively) for different $m_Q$ and $\Lambda$ values:
        m100 and m500 stands for the quirk mass value. L, 10L denotes
$\Lambda=(1,10)\ \Lambda_{QCD}$, 
        respectively.} 
\end{figure}

% Figure 2
\newpage
\vspace{2.0cm}
\begin{figure}
\includegraphics[width=10cm]{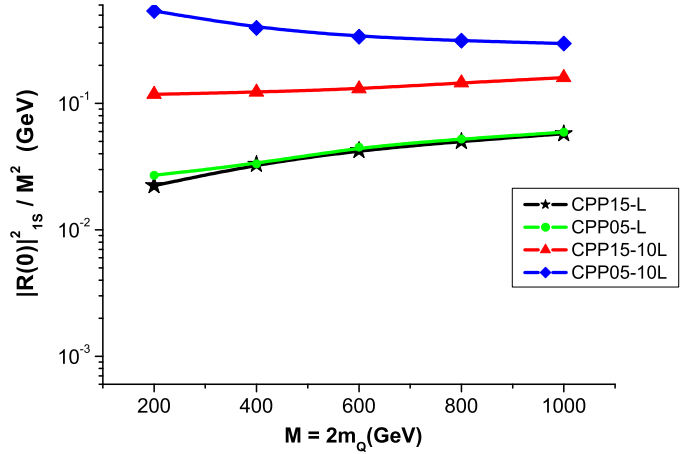}  
\caption{\label{fig:2} Values of $|R_{1S}(0)|^2/M^2$ (in $GeV$) 
        corresponding to the $1S$ level vs. $M = 2 m_Q$  (in $GeV$) using  CpP  potentials 
         with  $\nu=0.5; 1.5$  (labeled as CPP05, CPP15 respectively). 
         L,  10L denote $\Lambda=(1,10)\ \Lambda_{QCD}$, respectively.
         The curve corresponding to $\nu=1$ lies in between.} 
\end{figure}

\newpage

%\appendix
%\appendix\renewcommand{\theequation}{\thesection.\arabic{table}}
\section{Appendix: Tables}

\begin{table}[htb]
\caption{Mass level spacings with
respect to the ground state: $\Delta_{nl-1S}=M(nl)-M(1S)$ (MeV), 
using a Coulomb plus linear potential with
$m_Q=100$ GeV; $\Lambda=0.1 \cdot \Lambda_{QCD}= 25$ MeV;  
$|R_{nl}^{(l)}(0)|^2/M^{(2+2l)}$ (in GeV), and
mean square radius (in fm).} 
\label{table:1}
\newcommand{\m}{\hphantom{$-$}}
\newcommand{\cc}[1]{\multicolumn{1}{c}{#1}}
\renewcommand{\tabcolsep}{2pc} % enlarge column spacing
\renewcommand{\arraystretch}{1.2} % enlarge line spacing
\begin{tabular}{@{}cccc}
\hline
 $Q\bar{Q}$ LEVEL & $\Delta_{nl-1S}$ (MeV) & 
$|R_{nl}^{(l)}(0)|^2/M^{(2+2l)}$ & $\sqrt{\langle r^2\rangle}$\\
\hline
$(1S)$ & $0$   & $0.0092$& $0.06$\\
\hline
$(1P)$ & $153$   & $5.2\ 10^{-8}$ & $0.23$ \\
\hline
$(2S)$ & $154$ & $0.0012$     & $0.28$ \\
\hline
$(1D)$ & $183$  & $5.9\ 10^{-12}$     & $0.45$ \\
\hline
$(2P)$ & $184$  & $2.1\ 10^{-8}$     & $0.54$ \\
\hline
$(3S)$ & $185$  & $4.1\ 10^{-4} $    & $0.58$ \\
\hline
$(4S)$ & $198$  & $2.2\ 10^{-4} $    & $0.93$ \\
\hline
$(5S)$ & $206$  & $1.6\ 10^{-4} $    & $1.27$ \\
\hline
\end{tabular}\\[2pt]
 $\langle v^2 \rangle \approx 10^{-4}$\ 
\end{table}

\begin{table}[htb]
\caption{The same as in Table \ref{table:1} for a CpL potential with
$m_Q=100$ GeV; $\Lambda=0.4 \cdot \Lambda_{QCD}= 100$ MeV.} 
\label{table:2}
\newcommand{\m}{\hphantom{$-$}}
\newcommand{\cc}[1]{\multicolumn{1}{c}{#1}}
\renewcommand{\tabcolsep}{2pc} % enlarge column spacing
\renewcommand{\arraystretch}{1.2} % enlarge line spacing
\begin{tabular}{@{}cccc}
\hline
 $Q\bar{Q}$ LEVEL &  $\Delta_{nl-1S}$ (MeV) & $|R_{nl}^{l}(0)|^2/M^{(2+2l)}$ & $\sqrt{\langle r^2\rangle}$\\
\hline
$(1S)$ & $0$   & $0.016$ & $0.05$\\
\hline
$(1P)$ & $229$   & $1.7\ 10^{-7}$ & $0.17$ \\
\hline
$(2S)$ & $234$ & $0.0025$     & $0.21$ \\
\hline
$(1D)$ & $288$  & $8.3\ 10^{-11}$     & $0.28$ \\
\hline
$(2P)$ & $297$  & $1.1\ 10^{-7}$     & $0.34$ \\
\hline
$(3S)$ & $302$  & $0.0013$    & $0.37$ \\
\hline
$(4S)$ & $344$  & $9.5\ 10^{-4} $    & $0.52$ \\
\hline
$(5S)$ & $377$  & $7.9\ 10^{-4} $    & $0.65$ \\
\hline
\end{tabular}\\[2pt]
$\langle v^2 \rangle \approx [10^{-4},10^{-3}] $\ 
\end{table}

\begin{table}[htb]
\caption{The same as in Table \ref{table:1} for a CpL potential with
$m_Q=100$ GeV; $\Lambda=\Lambda_{QCD}= 250\ MeV$.} 
\label{table:3}
\newcommand{\m}{\hphantom{$-$}}
\newcommand{\cc}[1]{\multicolumn{1}{c}{#1}}
\renewcommand{\tabcolsep}{2pc} % enlarge column spacing
\renewcommand{\arraystretch}{1.2} % enlarge line spacing
\begin{tabular}{@{}cccc}
\hline
 $Q\bar{Q}$ LEVEL &  $\Delta_{nl-1S}$ (MeV) & $|R_{nl}^{l}(0)|^2/M^{(2+2l)}$ & $\sqrt{\langle r^2\rangle}$\\
\hline
$(1S)$ & $0$   & $0.025$ & $0.04$\\
\hline
$(1P)$ & $348$   & $6.6\ 10^{-7}$ & $0.12$ \\
\hline
$(2S)$ & $375$ & $0.0059$     & $0.15$ \\
\hline
$(1D)$ & $473$  & $7.0\ 10^{-10}$     & $0.18$ \\
\hline
$(2P)$ & $508$  & $6.0\ 10^{-7}$     & $0.21$ \\
\hline
$(3S)$ & $533$  & $0.0038$    & $0.24$ \\
\hline
$(4S)$ & $649$  & $0.0031$    & $0.32$ \\
\hline
$(5S)$ & $746$  & $0.0027$    & $0.38$ \\
\hline
\end{tabular}\\[2pt]
$\langle v^2 \rangle \approx 10^{-3} $\
\end{table}

\begin{table}[htb]
\caption{The same as Table \ref{table:1} for a CpL potential with 
$m_Q=100$ GeV; $\Lambda=\ 10 \cdot \Lambda_{QCD}= 2.5$ GeV.} 
\label{table:4}
\newcommand{\m}{\hphantom{$-$}}
\newcommand{\cc}[1]{\multicolumn{1}{c}{#1}}
\renewcommand{\tabcolsep}{2pc} % enlarge column spacing
\renewcommand{\arraystretch}{1.2} % enlarge line spacing
\begin{tabular}{@{}cccc}
\hline
 $Q\bar{Q}$ LEVEL &  $\Delta_{nl-1S}$ (GeV) & $|R_{nl}^{l}(0)|^2/M^{(2+2l)}$ & $\sqrt{\langle r^2\rangle}$\\
\hline
$(1S)$ & $0$   & $0.218$ & $0.02$\\
\hline
$(1P)$ & $2.51$   & $1.5\ 10^{-4}$ & $0.03$ \\
\hline
$(2S)$ & $3.42$ & $0.129$     & $0.04$ \\
\hline
$(1D)$ & $4.15$  & $ 3.1\ 10^{-7}$     & $0.045$ \\
\hline
$(2P)$ & $5.03$   & $2.1\ 10^{-4}$ & $0.05$ \\
\hline
$(3S)$ & $5.84$  & $0.109$    & $0.06$ \\
\hline
$(4S)$ & $7.88$  & $0.099$    & $0.08$ \\
\hline
$(5S)$ & $9.70$  & $0.092$    & $0.09$ \\
\hline
\end{tabular}\\[2pt]
$\langle v^2 \rangle \approx 10^{-2}$\ 
\end{table}

\begin{table}[htb]
\caption{The same as in Table \ref{table:1} for a CpL potential 
with $m_Q=500$ GeV; $\Lambda=\
\Lambda_{QCD}= 250$  MeV.}
\label{table:5}
\newcommand{\m}{\hphantom{$-$}}
\newcommand{\cc}[1]{\multicolumn{1}{c}{#1}}
\renewcommand{\tabcolsep}{2pc} % enlarge column spacing
\renewcommand{\arraystretch}{1.2} % enlarge line spacing
\begin{tabular}{@{}cccc}
\hline
 $Q\bar{Q}$ LEVEL &  $\Delta_{nl-1S}$ (MeV) & $|R_{nl}^{l}(0)|^2/M^{(2+2l)}$ & $\sqrt{\langle r^2\rangle}$\\
\hline
$(1S)$ & $0$   & $0.058$ & $0.008$\\
\hline
$(1P)$ & $831$   & $4.5\ 10^{-7}$ & $0.04$ \\
\hline
$(2S)$ & $850$ & $0.0063$     & $0.05$ \\
\hline
$(1D)$ & $1028$  & $2.0\ 10^{-11}$     & $0.07$ \\
\hline
$(2P)$ & $1042$  & $2.3\ 10^{-7}$     & $0.09$ \\
\hline
$(3S)$ & $1054$  & $0.0026$    & $0.1$ \\
\hline
$(4S)$ & $1162$  & $0.0017$    & $0.15$ \\
\hline
$(5S)$ & $1239$  & $0.0013$    & $0.19$ \\
\hline
\end{tabular}\\[2pt]
$\langle v^2 \rangle \approx 10^{-4} $\ 
\end{table}

\begin{table}[htb]
\caption{The same as in Table \ref{table:1} for a CpL potential with $m_Q=500$ GeV,  
$\Lambda=\ 10 \cdot \Lambda_{QCD}= 2.5$ GeV.} 
\label{table:6}
\newcommand{\m}{\hphantom{$-$}}
\newcommand{\cc}[1]{\multicolumn{1}{c}{#1}}
\renewcommand{\tabcolsep}{2pc} % enlarge column spacing
\renewcommand{\arraystretch}{1.2} % enlarge line spacing
\begin{tabular}{@{}cccc}
\hline
 $Q\bar{Q}$ LEVEL &  $\Delta_{nl-1S}$ (GeV) & $|R_{nl}^{l}(0)|^2/M^{(2+2l)}$ & $\sqrt{\langle r^2\rangle}$\\
\hline
$(1S)$ & $0$   & $0.187$ & $0.007$\\
\hline
$(1P)$ & $2.44$   & $1.0\ 10^{-5}$ & $0.017$ \\
\hline
$(2S)$ & $2.98$   & $0.048$ & $0.022$\\
\hline
$(1D)$ & $3.71$  & $2.0\  10^{-9}$     & $0.024$ \\
\hline
$(2P)$ & $4.18$  & $2.0\ 10^{-5}$     & $0.029$ \\
\hline
$(3S)$ & $4.64$  & $0.0350 $    & $0.033$ \\
\hline
$(4S)$ & $5.96$  & $0.030$    & $0.042$ \\
\hline
$(5S)$ & $7.11$  & $0.027$    & $0.051$ \\
\hline
\end{tabular}\\[2pt]
$\langle v^2 \rangle \approx 10^{-3}$\
\end{table}

\begin{table}[htb]
\caption{The same as in table \ref{table:1} for Coulomb plus power law potential with 
$\nu = 0.5$, $m_Q=100$ GeV; $\Lambda=\ \Lambda_{QCD}= 250$  MeV.} 
\label{table:7}
\newcommand{\m}{\hphantom{$-$}}
\newcommand{\cc}[1]{\multicolumn{1}{c}{#1}}
\renewcommand{\tabcolsep}{2pc} % enlarge column spacing
\renewcommand{\arraystretch}{1.2} % enlarge line spacing
\begin{tabular}{@{}cccc}
\hline
 $Q\bar{Q}$ LEVEL &  $\Delta_{nl-1S}$ (MeV) & $|R_{nl}^{l}(0)|^2/M^{(2+2l)}$ & $\sqrt{\langle r^2\rangle}$\\
\hline
$(1S)$ & $0$   & $0.027$ & $0.05$\\
\hline
$(1P)$ & $400$   & $9.6\ 10^{-7}$ & $0.11$ \\
\hline
$(2S)$ & $434$ & $0.0068$     & $0.14$ \\
\hline
$(1D)$ & $545$  & $6.7\ 10^{-11}$     & $0.17$ \\
\hline
$(2P)$ & $579$  & $7.6\ 10^{-7}$     & $0.21$ \\
\hline
$(3S)$ & $604$  & $0.0039$    & $0.23$ \\
\hline
$(4S)$ & $717$  & $0.0029$    & $0.32$ \\
\hline
$(5S)$ & $803$  & $0.0023$    & $0.41$ \\
\hline
\end{tabular}\\[2pt]
$\langle v^2 \rangle \approx 10^{-3} $\ 
\end{table}

\begin{table}[htb]
\caption{The same as in Table \ref{table:1} for a CpP potential
with $\nu = 0.5$, $m_Q=500$ GeV; 
$\Lambda=\ \Lambda_{QCD}= 250$ MeV.} 
\label{table:8}
\newcommand{\m}{\hphantom{$-$}}
\newcommand{\cc}[1]{\multicolumn{1}{c}{#1}}
\renewcommand{\tabcolsep}{2pc} % enlarge column spacing
\renewcommand{\arraystretch}{1.2} % enlarge line spacing
\begin{tabular}{@{}cccc}
\hline
 $Q\bar{Q}$ LEVEL &  $\Delta_{nl-1S}$ (MeV) & $|R_{nl}^{l}(0)|^2/M^{(2+2l)}$ & $\sqrt{\langle r^2\rangle}$\\
\hline
$(1S)$ & $0$   & $0.060$ & $0.01$\\
\hline
$(1P)$ & $880$   & $5.6\ 10^{-7}$ & $0.04$ \\
\hline
$(2S)$ & $911$ & $0.0072$     & $0.05$ \\
\hline
$(1D)$ & $1112$  & $6.0\ 10^{-12}$     & $0.07$ \\
\hline
$(2P)$ & $1132$  & $3.0\ 10^{-7}$     & $0.08$ \\
\hline
$(3S)$ & $1150$  & $0.0031$    & $0.09$ \\
\hline
$(4S)$ & $1276$  & $0.0020$    & $0.14$ \\
\hline
$(5S)$ & $1363$  & $0.0015$    & $0.18$ \\
\hline
\end{tabular}\\[2pt]
$\langle v^2 \rangle \approx 10^{-4} $\ 
\end{table}

\begin{table}[htb]
\caption{The same as in Table \ref{table:1} for a CpP potential with $\nu = 0.5$, 
$m_Q=100$ GeV; $\Lambda=\ 10 \cdot \Lambda_{QCD}= 2.5$  GeV.} \label{table:9}
\newcommand{\m}{\hphantom{$-$}}
\newcommand{\cc}[1]{\multicolumn{1}{c}{#1}}
\renewcommand{\tabcolsep}{2pc} % enlarge column spacing
\renewcommand{\arraystretch}{1.2} % enlarge line spacing
\begin{tabular}{@{}cccc}
\hline
 $Q\bar{Q}$ LEVEL &  $\Delta_{nl-1S}$ (GeV) & $|R_{nl}^{l}(0)|^2/M^{(2+2l)}$ & $\sqrt{\langle r^2\rangle}$\\
\hline
$(1S)$ & $0$   & $0.539$ & $0.014$\\
\hline
$(1P)$ & $5.09$   & $6.3\ 10^{-4}$ & $0.024$ \\
\hline
$(2S)$ & $6.94$ & $0.298$     & $0.031$ \\
\hline
$(1D)$ & $8.36$  & $3.1\ 10^{-6}$     & $0.032$ \\
\hline
$(2P)$ & $9.84$  & $0.0013$     & $0.040$ \\
\hline
$(3S)$ & $11.28$  & $0.227$    & $0.046$ \\
\hline
$(4S)$ & $14.63$  & $0.190$    & $0.060$ \\
\hline
$(5S)$ & $17.42$  & $0.167$    & $0.073$ \\
\hline
\end{tabular}\\[2pt]
$\langle v^2 \rangle \approx 10^{-2} $\
\end{table}

\begin{table}[htb]
\caption{The same as in Table \ref{table:1} for a CpP potential with $\nu = 0.5$, 
$m_Q=500$ GeV; $\Lambda=\ 10 \cdot \Lambda_{QCD}= 2.5$ GeV.} 
\label{table:10}
\newcommand{\m}{\hphantom{$-$}}
\newcommand{\cc}[1]{\multicolumn{1}{c}{#1}}
\renewcommand{\tabcolsep}{2pc} % enlarge column spacing
\renewcommand{\arraystretch}{1.2} % enlarge line spacing
\begin{tabular}{@{}cccc}
\hline
 $Q\bar{Q}$ LEVEL &  $\Delta_{nl-1S}$ (GeV) & $|R_{nl}^{l}(0)|^2/M^{(2+2l)}$ & $\sqrt{\langle r^2\rangle}$\\
\hline
$(1S)$ & $0$   & $0.298$ & $0.007$\\
\hline
$(1P)$ & $4.60$   & $9.0\ 10^{-5}$ & $0.012$ \\
\hline
$(2S)$ & $6.14$ & $0.107$     & $0.015$ \\
\hline
$(1D)$ & $7.39$  & $5.0\ 10^{-8}$     & $0.016$ \\
\hline
$(2P)$ & $8.44$  & $1.0\ 10^{-4}$     & $0.020$ \\
\hline
$(3S)$ & $9.61$  & $0.075$    & $0.023$ \\
\hline
$(4S)$ & $12.21$  & $0.061$    & $0.031$ \\
\hline
$(5S)$ & $14.33$  & $0.052$    & $0.038$ \\
\hline
\end{tabular}\\[2pt]
$\langle v^2 \rangle \approx 10^{-3} $\ 
\end{table}

\begin{table}[htb]
\caption{The same as in Table \ref{table:1} for a CpP potential with $\nu = 1.5$, 
$m_Q=100$ GeV; $\Lambda=\ \Lambda_{QCD}= 250$ MeV.} \label{table:11}
\newcommand{\m}{\hphantom{$-$}}
\newcommand{\cc}[1]{\multicolumn{1}{c}{#1}}
\renewcommand{\tabcolsep}{2pc} % enlarge column spacing
\renewcommand{\arraystretch}{1.2} % enlarge line spacing
\begin{tabular}{@{}cccc}
\hline
 $Q\bar{Q}$ LEVEL &  $\Delta_{nl-1S}$ (MeV) & $|R_{nl}^{l}(0)|^2/M^{(2+2l)}$ & $\sqrt{\langle r^2\rangle}$\\
\hline
$(1S)$ & $0$   & $0.022$ & $0.05$\\
\hline
$(1P)$ & $313$   & $4.6\ 10^{-7}$ & $0.13$ \\
\hline
$(2S)$ & $329$ & $0.0049$ & $0.16$ \\
\hline
$(1D)$ & $416$  & $1.5\ 10^{-11}$     & $0.20$ \\
\hline
$(2P)$ & $445$  & $4.4\ 10^{-7}$     & $0.23$ \\
\hline
$(3S)$ & $464$  & $0.0034$    & $0.25$ \\
\hline
$(4S)$ & $568$  & $0.0029$    & $0.33$ \\
\hline
$(5S)$ & $662$  & $0.0027$    & $0.39$ \\
\hline
\end{tabular}\\[2pt]
$\langle v^2 \rangle \approx 10^{-3} $\ 
\end{table}

\begin{table}[htb]
\caption{The same as in Table \ref{table:1} for a CpP potential $\nu = 1.5$,  
$m_Q=500$ GeV; $\Lambda=\ \Lambda_{QCD}= 250$ MeV.}
\label{table:12}
\newcommand{\m}{\hphantom{$-$}}
\newcommand{\cc}[1]{\multicolumn{1}{c}{#1}}
\renewcommand{\tabcolsep}{2pc} % enlarge column spacing
\renewcommand{\arraystretch}{1.2} % enlarge line spacing
\begin{tabular}{@{}cccc}
\hline
 $Q\bar{Q}$ LEVEL &  $\Delta_{nl-1S}$ (MeV) & $|R_{nl}^{l}(0)|^2/M^{(2+2l)}$ & $\sqrt{\langle r^2\rangle}$\\
\hline
$(1S)$ & $0$   & $0.058$ & $0.01$\\
\hline
$(1P)$ & $813$   & $3.7\ 10^{-7}$ & $0.04$ \\
\hline
$(2S)$ & $827$ & $0.0058$     & $0.05$ \\
\hline
$(1D)$ & $991$  & $2.0\ 10^{-12}$     & $0.08$ \\
\hline
$(2P)$ & $998$  & $1.8\ 10^{-7}$     & $0.10$ \\
\hline
$(3S)$ & $1005$  & $0.0022$    & $0.11$ \\
\hline
$(4S)$ & $1094$  & $0.0015$    & $0.16$ \\
\hline
$(5S)$ & $1198$  & $0.0011$    & $0.20$ \\
\hline
\end{tabular}\\[2pt]
$\langle v^2 \rangle \approx 10^{-4} $\
\end{table}

\begin{table}[htb]
\caption{The same as in Table \ref{table:1}  for a CpP potential 
with $\nu = 1.5$, $m_Q=100$ GeV; $\Lambda=\ 10 \cdot \Lambda_{QCD}= 2.5$ GeV.} 
\label{table:13}
\newcommand{\m}{\hphantom{$-$}}
\newcommand{\cc}[1]{\multicolumn{1}{c}{#1}}
\renewcommand{\tabcolsep}{2pc} % enlarge column spacing
\renewcommand{\arraystretch}{1.2} % enlarge line spacing
\begin{tabular}{@{}cccc}
\hline
 $Q\bar{Q}$ LEVEL &  $\Delta_{nl-1S}$ (GeV) & $|R_{nl}^{l}(0)|^2/M^{(2+2l)}$ & $\sqrt{\langle r^2\rangle}$\\
\hline
$(1S)$ & $0$   & $0.118$ & $0.027$\\
\hline
$(1P)$ & $1.12$   & $4.1\ 10^{-5}$ & $0.041$ \\
\hline
$(2S)$ & $1.88$ & $0.067$     & $0.057$ \\
\hline
$(1D)$ & $2.33$  & $3.1\ 10^{-8}$     & $0.060$ \\
\hline
$(2P)$ & $2.83$  & $5.8\ 10^{-5}$     & $0.070$ \\
\hline
$(3S)$ & $3.27$  & $0.059$    & $0.079$ \\
\hline
$(4S)$ & $4.52$  & $0.056$    & $0.097$ \\
\hline
$(5S)$ & $5.69$  & $0.054$    & $0.112$ \\
\hline
\end{tabular}\\[2pt]
$\langle v^2 \rangle \approx 10^{-2} $\ 
\end{table}

\begin{table}[htb]
\caption{The same as in Table \ref{table:1} for a CpP potential with
$\nu = 1.5$, $m_Q=500$ GeV; $\Lambda=\ 10 \cdot \Lambda_{QCD}= 2.5$ GeV.}
\label{table:14}
\newcommand{\m}{\hphantom{$-$}}
\newcommand{\cc}[1]{\multicolumn{1}{c}{#1}}
\renewcommand{\tabcolsep}{2pc} % enlarge column spacing
\renewcommand{\arraystretch}{1.2} % enlarge line spacing
\begin{tabular}{@{}cccc}
\hline
 $Q\bar{Q}$ LEVEL &  $\Delta_{nl-1S}$ (GeV) & $|R_{nl}^{l}(0)|^2/M^{(2+2l)}$ & $\sqrt{\langle r^2\rangle}$\\
\hline
$(1S)$ & $0$   & $0.160$ & $0.008$\\
\hline
$(1P)$ & $1.73$   & $5.0\ 10^{-6}$ & $0.023$ \\
\hline
$(2S)$ & $1.91$ & $0.026$     & $0.029$ \\
\hline
$(1D)$ & $2.04$  & $8.0\ 10^{-10}$     & $0.031$ \\
\hline
$(2P)$ & $2.62$  & $5.0\ 10^{-6}$     & $0.040$ \\
\hline
$(3S)$ & $2.80$  & $0.019$    & $0.044$ \\
\hline
$(4S)$ & $3.51$  & $0.016$    & $0.057$ \\
\hline
$(5S)$ & $4.15$  & $0.015$    & $0.067$ \\
\hline
\end{tabular}\\[2pt]
$\langle v^2 \rangle \approx 10^{-3} $\ 
\end{table}

\begin{table}[htb]
\caption{Mass level spacing  $\Delta_{2S-1S}=M(2S)-M(1S)$ (with $m_Q = 500$ GeV) using  
$\Lambda = (10, 20, 40) \cdot \Lambda_{QCD}$ and different $\bar{\alpha'}$  values.}
\label{table:15}
\newcommand{\m}{\hphantom{$-$}}
\newcommand{\cc}[1]{\multicolumn{1}{c}{#1}}
\renewcommand{\tabcolsep}{2pc} % enlarge column spacing
\renewcommand{\arraystretch}{1.2} % enlarge line spacing
\begin{tabular}{@{}ccc}
\hline
$\Lambda$ & $\bar{\alpha'}$ & $\Delta_{2S-1S}$ (GeV) \\
\hline
$10 \cdot \Lambda_{QCD}$ & $0.3$   & $\approx 10$ \\
\hline
$20 \cdot \Lambda_{QCD}$ & $0.3$   & $\approx 12$ \\
\hline
$40 \cdot \Lambda_{QCD}$ & $0.6$   & $\approx 30$ \\
\hline
\end{tabular}\\[2pt]
\end{table}

\begin{table}[htb]
\caption{Branching Ratios of $\psi_{Q\bar{Q}}(1S)\ (M=\ 2m_Q=\ 200\ GeV)$ and  $\psi_{Q\bar{Q}}(1S)\ (M=1000\  GeV)$  to $SM$ quarks ($q\bar{q}$),leptons ($l\bar{l}$), and other boson decays. $\Gamma$(CpL,CpP15,CpP05) stands for the total decay width (in $KeV$) using each potential. }
\label{table:16}
\newcommand{\m}{\hphantom{$-$}}
\newcommand{\cc}[1]{\multicolumn{1}{c}{#1}}
\renewcommand{\tabcolsep}{0.2pc} % enlarge column spacing
\renewcommand{\arraystretch}{1.2} % enlarge line spacing
\begin{tabular}{@{}ccccc}
\hline
$\psi_{Q\bar{Q}}(1S)\ (M=200\ GeV)$& & & & \\
\hline
$\Lambda (MeV)$ & $\Gamma$(CpL,CpP15,CpP05)$(KeV)$& $BR(l\bar{l})(\%)$& $BR(q\bar{q})(\%)$& $BR(Boson)(\%)$ \\
\hline
$25 $  & $( 31, --, -- )$   & $ 17$& $76$& $7$\\
\hline
$100 $  & $( 54, --, -- )$   & $ 16$& $76$& $8$\\
\hline
$250 $  & $( 86,\  78,\  94 )$   & $ 16$& $75$& $9$\\
\hline
$2500$  & $(833, 451, 2060 )$   & $15$& $68$& $17$\\
\hline
$\psi_{Q\bar{Q}}(1S)\ (M=1000\ GeV)$& & & & \\
\hline
$\Lambda (MeV)$ & $\Gamma$(CpL,CpP15,CpP05)$(KeV)$& $BR(l\bar{l})(\%)$& $BR(q\bar{q})(\%)$& $BR(Boson)(\%)$ \\
\hline
$250 $  & $(165, 164, 169)$   & $ 16$& $80$& $4$\\
\hline
$2500$  & $(555, 474, 882)$   & $16$& $76$& $8$\\
\hline
\end{tabular}\\[2pt]
\end{table}

\end{document}